\documentclass[a4paper]{article}

\usepackage{INTERSPEECH2021}
\usepackage[T2A]{fontenc}
\usepackage{caption}
\usepackage{subcaption}

\title{Golos: Russian Dataset for Speech Research}
\name{Nikolay Karpov, Alexander Denisenko, Fedor Minkin}

\address{Sber, Russia}
\email{karpnv@gmail.com, alexander.denisenko@phystech.edu, minkin.f.a@sberbank.ru}

\begin{document}

\maketitle
\begin{abstract}
 
This paper introduces a novel Russian speech dataset called Golos, a large corpus suitable for speech research. The dataset mainly consists of recorded audio files manually annotated on the crowd-sourcing platform. The total duration of the audio is about 1240 hours. We have made the corpus freely available to download, along with the acoustic model with CTC loss prepared on this corpus. Additionally, transfer learning was applied to improve the performance of the acoustic model. In order to evaluate the quality of the dataset with the beam-search algorithm, we have built a 3-gram language model on the open Common Crawl dataset. The total word error rate (WER) metrics turned out to be about 3.3\% and 11.5\%.

\end{abstract}
\noindent\textbf{Index Terms}: speech recognition, open dataset, Russian language, speech corpus, acoustic model, language model

\section{Introduction}
We believe that open data is one of the key drivers of the recent success in the field of artificial intelligence. In particular, automatic speech recognition (ASR) algorithms became much better in quality and more robust during the recent years. These new algorithms allow researchers to create conversational systems with good user experience; as a result, such technologies become more popular. That disrupts traditional business strategies and provides a foundation for many new business ideas and benefits for innovators. 
  
Despite the existence of outstanding initiatives such as MLS dataset \cite{pratap2020mls}, there is a lack of manually annotated large scale speech corpora in Russian that would be freely available and suitable for training and testing speech recognition systems.
 
This article is dedicated to present our new open Russian speech dataset with manual annotation. It can be useful for many research projects such as \cite{savchenko2008analyse, gubochkin2013cl} which need labeled audio data. We provide an example on how to train an acoustic model on this data using the open source NeMo toolkit \cite{kuchaiev2019nemo}. We also demonstrate an improvement in performance using pre-trained English acoustic models and quality benefits from the language model. The highlights of this paper are the following:
\begin{enumerate}
\item An open audio corpus with 1240 hours of manually annotated speech in Russian.\footnote{https://github.com/sberdevices/golos}
\item An example of an acoustic model trained on our corpus.
\item The empirical results of transfer learning for the Russian acoustic model using the pre-trained English one.
\item An evaluation of the acoustic model using a beam-search decoder with the language model trained on an open dataset.
\end{enumerate}

In Section 2 we review the related work that inspired us to share our corpus. In Section 3 we describe the pipeline that we used to build the corpus. Section 4 describes the structure of the dataset. Section 5 presents the acoustic model trained on this dataset and the experimental results of transfer learning. Finally, in Section 6 we provide the acoustic model evaluation together with the language model, which we also made available.

\section{Related Work}

There are few large scale data collections for speech recognition available now. 

\subsection{Mono-lingual ASR datasets}

Let us mention only several mono-lingual datasets which we think are the most important in the field of Russian speech recognition.

First one is LibriSpeech \cite{panayotov2015librispeech} which includes about 1000 hours of English audio-books. It is derived from the big LibriVox dataset and distributed under an open license. Second one is an audio part of the Wall Street Journal (WSJ) corpus \cite{paul1992design} which contains about 400 hrs. of speech data. These two datasets are usually used as a public benchmarks to compare new algorithms. It helps to push forward the state of the art (SoTA) quality level of speech recognition.

Open-STT is the only Russian language large-scale dataset. It consists of more than 15 000 hours of audio \cite{veysov2020towardimagenetstt}. Unfortunately, its annotation is not manually created. Transcripts are derived by doing alignment or using the ASR system. 

\begin{table}[th]
  \caption{Content of Open STT Russian corpus}
  \label{tab:openstt}
  \centering
  \begin{tabular}{ llcr }
    \toprule
    Domain & Annotation & Utterances & Hours~~~  \\
    \midrule
    Radio &     Alignment & 8.3M & 11 996 ~~~  \\
    Public Speech & Alignment  & 1,7M & 2 709 ~~~ \\
    YouTube & Subtitles  & 2,6M & 2 117 ~~~ \\
    Audiobooks & Alignment  & 1,3M & 1 632 ~~~ \\
    Calls & ASR  & 695K & 819 ~~~ \\
    Other & TTS  & 1.9M & 835 ~~~ \\
    \bottomrule
  \end{tabular}
\end{table}

\subsection{Multi-lingual ASR datasets}

We mainly focus on multi-lingual corpuses which include Russian data. VoxForge\footnote{http://www.voxforge.org} may be the oldest and still available open speech resource. It remains low-scale, and includes only about 300 hours in total, and even less in Russian.

Common Voice \cite{ardila2019common}, is a scalable solution with more than 30 languages available, including Russian. It keeps growing, with 4500 (validated) hours currently available. There are only 111 hours of Russian labeled data.

LibriVox\footnote{https://librivox.org} includes nearly 80 506 hours in total, but only 172 hours of Russian audiobooks read by volunteers from all around the world.

The M-AILABS\footnote{https://www.caito.de/2019/01/the-m-ailabs-speech-dataset} speech dataset includes the data based on LibriVox and Project Gutenberg\footnote{http://www.gutenberg.org}. The data consists of nearly a thousand hours of audio in total and 47 hours of Russian speech.

MLS \cite{pratap2020mls} is a large-scale multilingual dataset for speech research, which, like the LibriSpeech and M-AILABS, is derived from the LibriVox dataset, and consists of 8 languages, including about 44.5K hours of English and a total of about 6K hours of other languages. However, it doesn't contain any Russian language data.

\section{Data Collection Pipeline}

This section describes the main steps during data collection and preparation. The pipeline is developed to solve the cold start problem when we have a new voice powered product but there is no real user data yet. It includes four steps.

\subsection{Templates Creation}
We create a list of domains which is suitable for our products: music, films, organizations, names, addresses, etc.
For each domain we develop so-called templates - structures which let us create highly plausible text queries within a certain domain. Basically, they describe the way in which different entities (such as commands, movie titles, actor names and specific forms of these entities - including variations in case and gender) might be arranged so that the resulting query might have been a result of some actual use-case.

For example the template "command + film" can be instantiated as "Put on Terminator 1"

\subsection{Audio Generation}
Using such templates we consistently generate tens and hundreds of thousands of text queries. We then proceed to voicing them with real people's voices. We use two types of voice sources: the popular crowd-sourcing platform Yandex.Toloka\footnote{https://toloka.yandex.ru} and studio recordings through our smart screen called SberPortal.

\subsection{Crowd-sourced Validation}
However, we cannot be certain that all the labellers have said exactly what was written in their assignment. We neither want to train on such data, nor test or validate on it, so we filter them out. We do so by presenting each pair of text with the crowdsourced audio-file to 5 people on the very same labelling platform and ask them whether or not the audio matches the text in front of them. If 5 out of 5 people say that it does, then we consider the text and the audio to match each other. If less than 5 people say so - we don't use this pair.

\subsection{Assisted Transcribation}
We want our corpus to be monolingual and only consist of Cyrillic letters and spaces. However, there are many sentences containing Latin and diacritic letters, numbers, special symbols, etc. In order to get rid of such symbols but preserve the data - we use Yandex.Toloka again. Labellers are being presented with the original text which might contain any kind of symbols and the corresponding audio which was validated in the previous step. The labellers are asked to write down the text using only Cyrillic symbols and spaces. 

It turned out to be vital for us to provide not only audios, but also hints in the form of original text, as it might be extremely difficult to transcribe the audio of a Russian who is pronouncing "Annenmaykantereit Barfuß am Klavier"  without any assistance. We presented each pair of audio and text to 5 labellers. We took a look at the most common transcription in Cyrillic - and if there were at least two people who have written the same transcription - we use it as a true label when training our acoustic model.

\section{Dataset Overview}

There are two types of audio sources in our corpus. The first one is a crowd-sourcing platform called Yandex.Toloka. Further in text we call it “Crowd domain”. The second one is our main source of audio input - the smart screen “SberPortal”. The development stage of SberPortal has started before its sales commenced. So we had to generate audio through SberPortal in the studio by emulating the real user environment. To do so we used 1 meter, 3 meters and 5 meters distances between the speaker and the device. Further in text we call it a “Farfield domain” because of the distance which is usually quite large.

Each of the sources we split into train and test subsets. The testing part of Crowd set consists of about 10 000 files, Farfield test set is about 2 000. The exact dataset separation is shown in the Table~\ref{tab:dataset_separation}. 

\begin{table}[th]
  \caption{Golos Dataset Content Separation.}
  \label{tab:dataset_separation}
  \centering
  \begin{tabular}{ l|r|r|r|r }
    \toprule
    \multicolumn{1}{l|}{\textbf{Domain}} & \multicolumn{2}{c|}{\textbf{Train files and hours}} & \multicolumn{2}{c}{\textbf{Test files and hours}} \\
    \midrule
    Crowd  & 979796 & 1095~~~h.  & 9994 & 11.2 h. \\
    Farfield & 124003 & 132.4 h. &  1916 & 1.4 h. \\
    \bottomrule
    Total  & 1103799 & 1227.4 h. & 11910 & 12.6 h. \\
  \end{tabular}
\end{table}

We don't use any personal information about the speakers such as age, gender, user id. The recordings are anonymized and the separation is done without it. The voice of the same user could appear both in the train and test subsets. 

\begin{table}[t]
  \caption{Golos Training Set Description.}
  \label{tab:dataset_description}
  \centering
  \begin{tabular}{lrr}
    \toprule
    \textbf{Description}      & \textbf{Crowd  subset}     & \textbf{Farfield subset} \\
    \midrule
    Count & 979796 files & 124003 files \\
    Mean & 4.0~~ sec. & 3.8 ~~ sec. \\
    STD &  1.9~~ sec. & 1.6 ~~ sec. \\
    Min & 0.02 sec. & 0.002 sec. \\
    50th percentile& 3.7~~ sec. & 3.5 ~~ sec. \\
    95th percentile& 7.3~~ sec. & 6.8 ~~ sec. \\
    99th percentile& 10.5~~ sec. & 9.6 ~~ sec. \\
    Max & 56.3~~ sec. & 23.5 ~~ sec. \\
    \bottomrule
  \end{tabular}
\end{table}

\begin{figure}[ht]
\begin{subfigure}{.49\linewidth}
  \centering
  \includegraphics[width=1.\linewidth]{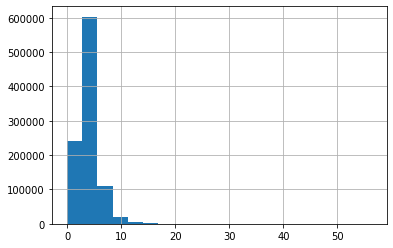}  
  \caption{Crowd domain.}
  \label{fig:sub-first}
\end{subfigure}
\begin{subfigure}{.49\linewidth}
  \centering
  \includegraphics[width=1.\linewidth]{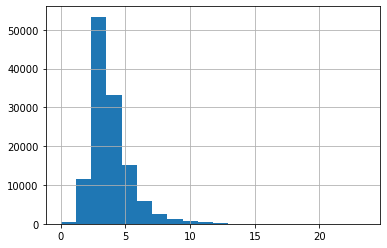}  
  \caption{Farfield domain.}
  \label{fig:sub-second}
\end{subfigure}
\caption{The number of samples depends on sample duration (sec.) in Golos training set.}
\label{fig:duration_distribution}
\end{figure}

Some statistical data like percentiles, mean and standard deviation (STD) of the duration values of the training sets is provided in the Table \ref{tab:dataset_description}. On the Figure \ref{fig:duration_distribution} a histogram is shown of the duration distribution of utterances in the training set.

To perform any limited or no supervision experiments with models we provide additional subsets of the training set with durations 100 hours, 10 hours, 1 hour, 10 minutes, as it was in the previous work \cite{kahn2020libri}.

\section{Acoustic Model}

The best way to show the quality of data is to train some model on it and to estimate the quality metrics of the model. We estimate the word error rate (WER) because it is a common performance metric of speech recognition systems.  

\subsection{Experiment Setup}

For an acoustic model we chose QuartzNet15x5 neural network \cite{kriman2020quartznet}. Its architecture is shown on the Table \ref{tabular:Quartznet}. The model starts with a convolutional layer $C_1$ followed by a sequence of 5 groups of blocks. Blocks in the group are identical, each block $B_k$ consists of R time-channel separable K-sized convolutional modules with C output channels. Each block is repeated S times. The model has 3 additional convolutional layers ($C_2, C_3, C_4$) at the end.

The training procedure is done by using the open source NeMo toolkit \cite{kuchaiev2019nemo}. We train our model on the Nvidia DGX-2 with 16 GPU cards Tesla v100 with a batch size of 88 per GPU, and accumulate gradients every 10 batches, so our effective batch size is $16 \cdot 88 \cdot 10 = 14080$. In order to decrease the memory footprint and training time, we used mixed-precision training \cite{micikevicius2017mixed}.

We trained an acoustic model on the randomly shuffled training part of the Golos dataset and evaluated it on two test sets separately (Crowd and Farfield domains). Data augmentation is carried out using SpecAugment \cite{park2019specaugment} without time warping deformation. We don't use dropouts during training. Training is conducted with the help of the NovoGrad \cite{ginsburg2019stochastic} optimizer ($\beta_1$ = 0.95, $\beta_2$ = 0.5) with a cosine annealing learning rate policy. About 5\% steps of the learning rate warm-up helps to stabilize early training with maximum learning rate of 0.01, and weight decay 0.001. Bigger learning rates lead to gradient overflow and infinite loss value in our experiments. 

\begin{table}[t]
  \caption{QuartzNet15x5 Architecture with Outputs for Russian Letters. }
  \label{tabular:Quartznet}
  \centering
  \begin{tabular}{ccccc}
    \toprule
    \textbf{Block}  & \textbf{R} & \textbf{K} & \textbf{C} & \textbf{S}     \\
    \midrule
    $C_1$  & 1 & 33 & 256 &  1  \\
    \midrule
    $B_1$  & 5 & 33 & 256 &  3  \\
    $B_2$  & 5 & 39 & 256 &  3  \\
    $B_3$  & 5 & 51 & 512 &  3  \\
    $B_4$  & 5 & 63 & 512 &  3  \\
    $B_5$  & 5 & 75 & 512 &  3  \\
    \midrule
    $C_2$  & 1 & 87 & 512 &  1  \\
    $C_3$  & 1 & 1 & 1024 &  1  \\
    $C_4$  & 1 & 1 & 34 &  1    \\
    \bottomrule
    \textbf{Params,M}  &  &  & &  18.9     \\
  \end{tabular}
\end{table}


\subsection{Transfer Learning}

Transfer learning is the key element in the recent success of neural networks, so it is widely used in the industry. For example, ImageNet dataset is often utilized for creating pre-trained models in computer vision, and in the natural language processing field pre-trained BERT models are usually used. 

The QuartzNet15x5 acoustic model pre-trained on 3300 hours of public data in English is publicly available. The Russian version of this model is shown on the Table \ref{tabular:Quartznet}. The only difference from the English version is in the last layer $C_4$, because the target language has a different alphabet. For Cyrillic alphabet, there are 34 possible outputs - 32 letters (excluding letter ё), whitespace and blank symbols. When training the model using the Latin alphabet, there are 29 possible outputs - 26 letters, whitespace, blank and apostrophe symbols. Our transfer learning pipeline is as follows:
\begin{itemize}
    \item Take layers weights $C_1, B_1, B_2, B_3, B_4, B_5, C_2, C_3$ from pre-trained network as is and initialize a new network with them.
    \item Map similar character (for instance "k"  to "к" ) weights from the old $C_4$ layer to the new one.
    \item Randomly initialize weights for new characters in the layer $C_4$
    \item Train the resulting network on the target dataset starting with the same learning rate that was used to pre-train the model from scratch.
\end{itemize}

\begin{figure}[ht]
\begin{subfigure}{.49\linewidth}
  \centering
  \includegraphics[width=1.\linewidth]{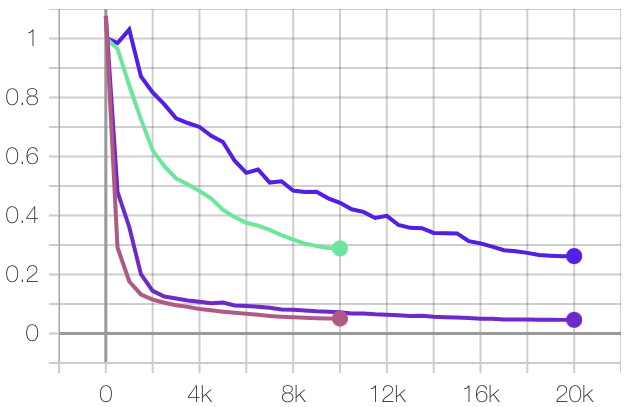}  
  \caption{Crowd domain.}
  \label{fig:crowd_domain}
\end{subfigure}
\begin{subfigure}{.49\linewidth}
  \centering
  \includegraphics[width=1.\linewidth]{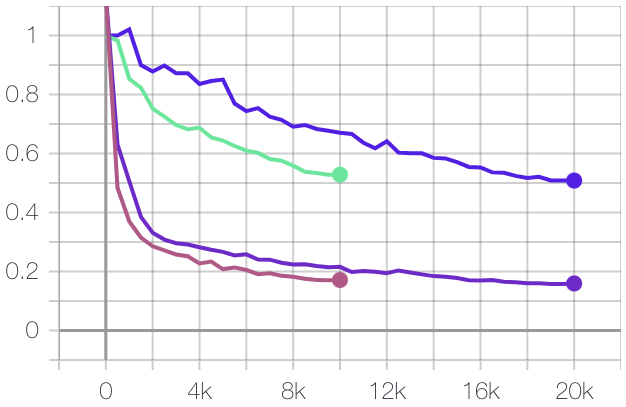}  
  \caption{Farfield domain.}
  \label{fig:portal_domain}
\end{subfigure}
\caption{Greedy WER dependence on transfer learning. Green - 10K from scratch, Blue - 20K from scratch. Red - 10K from En., Violet - 20K from En.}
\label{fig:portal_crowd_domain}
\end{figure}

Figures \ref{fig:crowd_domain} and \ref{fig:portal_domain} demonstrate how our transfer learning procedure affects greedy WER along the training process. There are four training setups: 10 000 steps with random initialization (Green - 10K from scratch), 10 000 steps with English initialization (Red - 10K from En.), 20 000 steps with random initialization (Blue - 20K from scratch), 20 000 steps with English initialization (Violet - 20K from En.). Each of the training setups has two evaluation sets - Crowd and Farfield - so we have eight curves. We can see that curves with random initialization are positioned much higher than initialised by the English model. It means that transfer learning always boosts our performance. WER of the models trained from scratch is big (more then 20\%) because number of training steps is quite small. If we train longer then the gap to the English initialisation becomes smaller but still remains.

Table \ref{tab:wer_50k} shows the final values of the greedy WER for our domains. There are three experiments with different training durations. 50 000 steps took about 8 days long, 20 000 steps took 2 days 21 hours, 10 000 steps took 1 day 19 hours. These WER scores were calculated using the greedy decoder. The best values are 4.327\% and 15.28\% on the Crowd and Farfield domain respectively. They were shown by the model which was trained for 50 000 steps.  

\begin{table}[t]
  \caption{Transfer Learning Influence on greedy WER \%.}
  \label{tab:wer_50k}
  \centering
  \begin{tabular}{lrr}
    \toprule
    \textbf{Training procedure } & \textbf{Crowd} & \textbf{Farfield} \\
    \midrule
    10K from scratch   &  28.84\%   & 52.82\% \\
    10K from En.    & 5.095\%       & 17.13\%  \\
    \midrule
    20K from scratch   &  26.24\%   & 50.82\% \\
    20K from En.    & 4.629\%       & 15.95\% \\
    \midrule
    50K from En.    & 4.327\%       & 15.28\% \\
    \bottomrule
  \end{tabular}
\end{table}

\begin{figure}[ht]
\begin{subfigure}{.49\linewidth}
  \centering
  \includegraphics[width=1.\linewidth]{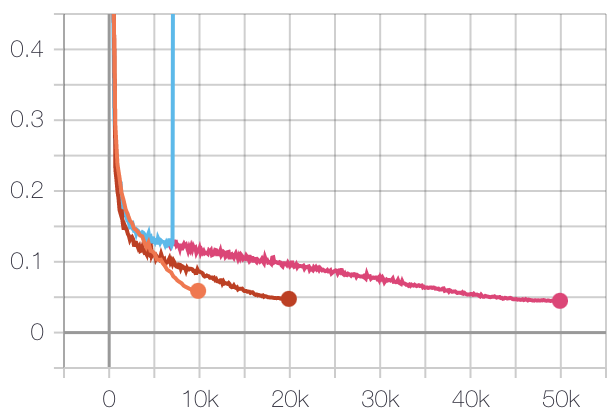}  
  \caption{Crowd domain.}
  \label{fig:wer_steps1}
\end{subfigure}
\begin{subfigure}{.49\linewidth}
  \centering
  \includegraphics[width=1.\linewidth]{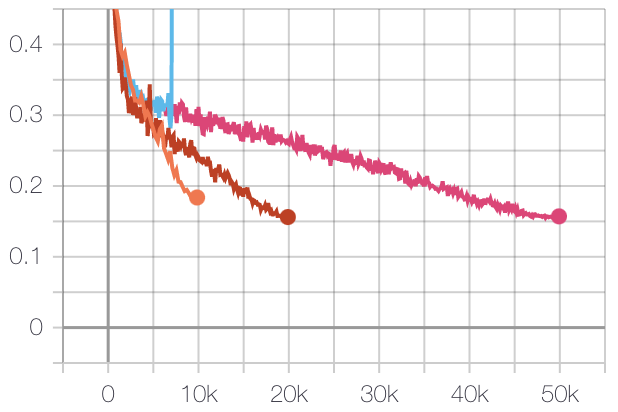}  
  \caption{Farfield domain.}
  \label{fig:wer_steps2}
\end{subfigure}

\caption{Greedy WER depends on the training duration with English initialisation. Orange - 10K, Red - 20K, Blue and Pink - 50K. }
\label{fig:wer_steps}
\end{figure}

\section{Language Model}

The last experiment was to train an acoustic model using Golos and Common Voice \cite{ardila2019common} datasets together and evaluate it with beam-search decoder and language model. Common Voice contains dev, test and train sets which duration is about 100 hours. Figure \ref{fig:wer_steps} demonstrates how greedy WER is changing along the whole training process for three training setups (10000, 20000, 50000 steps) initialized by the pre-trained English model.

We create a language model using the Russian part of the Common Crawl dataset\footnote{https://commoncrawl.org} and KenLM Language Model Toolkit \cite{heafield-2011-kenlm}. The Common Crawl is a repository of web crawl data that can be accessed and analyzed by anyone. KenLM toolkit allows us to create a very fast n-gram language model.

We have created three external 3-gram language models. The first one is using clean preprocessed texts from the Common Crawl dataset (CC LM). During preprocessing we had removed punctuation, space tokens and other extra symbols. The second one was built with the help of transcription of training set audios (train LM). The third model is a merge of these two 3-gram models with 50/50 percent weights (CC+train LM).

\begin{table}[t]
  \caption{Influence of Common Crawl (CC) Language Model on WER \% for Golos and Common Voice (CV) validation sets.}
  \label{tab:language_model}
  \centering
  \begin{tabular}{lrrrr}
    \toprule
    \textbf{Decoder} & \textbf{Crowd} & \textbf{Farfield} & CVdev & CVtest \\
    
    \midrule
    Greedy & 4.39\% & 14.95\% & 9.31\% & 11.28\% \\
    CC LM & 4.71\% & 12.5\% & 6.34\% & 7.98\% \\
    Train LM & 3.55\% & 12.38\% & - & - \\
    CC+train LM & 3.32\% & 11.49\% & 6.4\% & 8.06\%\\
    \bottomrule
  \end{tabular}
\end{table}

We use these created language models for an inference with beam-search decoder and following algorithm parameters: beam size=16, alpha=2, beta=1.5. Alpha is the amount of importance to place on the N-gram language model. Beta is a penalty term given to longer word sequences. Table \ref{tab:language_model} shows how the beam-search decoder with our three language models influences the resulting WER. The best WER values with the language model are 3.32\% and 11.49\%.

\section{Conclusions}

In this paper we reveal an open Russian language dataset. It is a large corpus suitable for speech research. It consists of audios obtained both from the crowd-sourcing platform and from the studio with far field settings. All 1240 hours of the audio are manually annotated.

Using our new corpus we have trained a QuartzNet acoustic model with CTC loss. The best performance of the acoustic-model was achieved with the help of transfer learning from a pre-trained model on English language. 

Additionally, we built a 3-gram language model on the open Common Crawl dataset and merged it with the train set transcriptions. Using a beam-search algorithm, our resulting model achieves 3.3\% and 11.5\% WER values on Crowd and Farfield datasets respectively.

All the data and models are freely available for downloading at the Github repository\footnote{https://github.com/sberdevices/golos}.



\bibliographystyle{IEEEtran}

\bibliography{mybib}

\end{document}